\documentclass[preprint,12pt]{elsarticle}

\usepackage{amssymb}

\journal{Physica B}

\begin{document}

\begin{frontmatter}



\title{Two-state Weiss model for the anomalous thermal expansion in 
EuNi${}_{2}$P${}_{2}$}


\author{Y. Kakehashi\corref{cor1}}
\author{S. Chandra}
\cortext[cor1]{Corresponding author. Tel: +81 098-895-8510; 
fax: +81 098-895-8509. E-mail address: yok@sci.u-ryukyu.ac.jp 
(Y. Kakehashi), to be published in Physica B}

\address{Department of Physics and Earth Sciences,
Faculty of Science, \\
University of the Ryukyus, \\
1 Senbaru, Nishihara, Okinawa, 903-0213, Japan}

\begin{abstract}
It has recently been found that the single crystalline 
EuNi${}_{2}$P${}_{2}$
shows an anormalous thermal expansion and its scaling relation to the
average 4$f$ electron number.  We point out that a phenomenological
2-state Weiss model can explain these behaviors.  We also show that 
the model leads to the observed effective Bohr magneton number per atom 
(7.4 $\mu_{\rm B}$) in the high-temperature limit, and is consistent 
with the temperature dependence of experimental susceptibility.  
Moreover the model predicts that the 4$f$ contribution to the thermal 
expansion coefficient is in proportional to that of the specific heat. 
The results suggest that the strong temperature dependence found 
in these experimental data originates in the excitations from the 
heavy-fermion ground state to the independent 4$f^{7}$ atomic states.
\end{abstract}

\begin{keyword}
Thermal expansion, Weiss model, 2$\gamma$-state model, 
EuNi${}_{2}$P${}_{2}$, Valence fluctuations, Heavy-fermions


\end{keyword}

\end{frontmatter}


\section{Introduction}
\label{}
Europium compounds are known to show a variety of magnetic and thermal
properties due to existence of different valence 
states~\cite{novik82,gupta87,kitagawa02,tsutsui09,takikawa10}.  In the
divalent compounds, Eu ions have 4$f^{7}$ configuration (Eu${}^{2+}$)
with atomic spin $S=7/2$, angular momentum $L=0$, and total angular
momentum $J=7/2$ showing a large magnetic moment, while
in the trivalent compounds, Eu ions have 4$f^{6}$ configuration 
(Eu${}^{3+}$) with 
$S=L=3$, and $J=0$, thus no magnetic moment.  Some of these compounds
such as EuPd${}_{2}$Si${}_{2}$~\cite{sampat81}, 
EuNi${}_{2}$(Si${}_{0.18}$Ge${}_{0.82}$)${}_{2}$~\cite{matsuda09}, and 
EuRh${}_{2}$Si${}_{2}$~\cite{mitsuda12}  show the
valence instability with increasing temperature, pressure, and magnetic
field.  In particular, EuNi${}_{2}$P${}_{2}$ has recently received much
attention  because it shows both heavy-fermion and mixed-valence
behaviors. In fact, M\"{o}ssbauer isomer-shift experiment reports a mixed
valence state even at zero temperature~\cite{nagara85} and 
the electronic specific heat coefficient $\gamma$ shows a large value 
of 100 mJ/(K${}^{2} \cdot$mol)~\cite{fisher95}. 

Quite recently, Hiranaka {\it et. al.}~\cite{hiranaka13} grew 
the single crystalline
EuNi${}_{2}$P${}_{2}$ and carried out systematic measurements of
resistivity, specific heat, susceptibility, and thermal expansion.  The
low-temperature data of resistivity and specific heat show the
heavy-fermion behavior with a large electronic specific heat coefficient 
$\gamma=93$ mJ/(K${}^{2} \cdot$ mol).  
The susceptibility data show the Curie-Weiss
behavior with the effective Bohr magneton number $p_{\rm eff}=7.4$
$\mu_{\rm B}/$Eu and the Weiss constant $\Theta=-120$ K in the high
temperature regime, and becomes 
almost constant ($\approx 0.04$ emu/mol) below 50 K.  Anomalous 4$f$ 
electronic contribution to the volume expansion $(\Delta V/V)_{4f}$
is found to increase rapidly up to 100 K and tends to saturate with 
increasing temperature.  The new aspect which they found is that the 
temperature variation of
$(\Delta V/V)_{4f}$ scales well to that of the average Eu valence.
Because the present system shows both the heavy-fermion and mixed-valence
behaviors and the anomalous volume expansion continues up to 200 K, 
which is far above the experimentally suggested Kondo temperature 
$T_{\rm K}$ ($\sim 80$ K), the scaling relation may be controlled by 
an extra-parameter other than $T_{\rm K}$, {\it i.e.}, a valence
fluctuation temperature $T_{v}$. 

In this paper, we point out that the temperature dependence of volume
thermal expansion and its scaling relation to the average 4$f$ electron
number found in experiment are basically explained by a simple and 
phenomenological 2$\gamma$-state model.  
The model was first proposed by Weiss~\cite{weiss63} to explain
the anomalous thermal expansion of $\gamma$Fe and the Fe-Ni Invar
alloys~\cite{wasser90}.  
It assumes the existence of two magnetic states, a low-spin
small-volume state and a high-spin  large-volume state.  The former is
assumed to be the ground state in $\gamma$Fe, while the latter is
assumed to be stabilized at the ground state in the Invar 
alloys~\cite{wasser90,schilf99,lawson06}.  One can describe the
excitations from the ground state to the independent excitations on
sites with use of the Weiss model even in the 
itinerant electron system. 

We remark that apart from the anomalous volume expansion, the Weiss
model is somewhat similar to the interconfiguration fluctuation (ICF)
model for some rare-earth
compounds~\cite{nagara85,sales75,franz80,wada96,paramanik13}. The
present model however differes from the ICF model in the following
points.
(1) The Weiss model assumes the nondegenerate metallic ground state
({\it i.e.}, a heavy-fermion state in the present case), while the ICF
model assume the 4f atomic state with an integral 4f electron number as
the ground state. (2) The Weiss model does not make use of any other
assumptions, while the ICF model introduces phenomenologically a
fluctuation temperature ($T_{sf}$) corresponding to the width of the 4f
level state. 

In the following section, we apply the concept of the 
2$\gamma$-state model in order to
explain a strong temperature dependence which is seen
in the experimental data of 
EuNi${}_{2}$P${}_{2}$.  
With use of the 2-state Weiss model, we will demonstrate that the model
explains an overall temperature dependence of thermal expansion, its
scaling relation to the 4$f$ electron number, the specific heat, and the
susceptibility.  We summalize the conclusion in the last section 3.

\section{Two-state Weiss model for EuNi${}_{2}$P${}_{2}$}

\subsection{Thermal expansion and scaling relation}

Experimentally, a heavy-fermion state is realized in
EuNi${}_{2}$P${}_{2}$ at low temperatures.
We therefore assume that the
ground state is a nonmagnetic heavy-fermion state with a small volume 
per atom $V_{\rm L}$ as found experimentally, and call this state `the
low spin state' hereafter.  We neglect the low energy excitations
associated with the heavy-fermions which yields the $T$-linear 
specific heat.  Instead, we take into account independent excitations to
the atomic Eu${}^{2+}$ state with 
$J=J_{\rm H} \, (=7/2)$ and a large volume $V_{\rm H}$.
We call the single-site excited state on each site 
`the high spin state'.
The volume per Eu atom is then characterized by $J$ as $V(J)$, and 
the thermal average $V$ is given by 
\begin{eqnarray}
V = \frac{\displaystyle \sum_{J}\sum_{M=-J}^{J} V(J) \, e^{-\beta E_{J}}}
{\displaystyle \sum_{J}\sum_{M=-J}^{J} e^{-\beta E_{J}}} \, .
\label{vol0}
\end{eqnarray}
Here $\beta$ is the inverse temperature and $E_{J}$ denotes the energy
of the system with the state $J$.  
Note that we allocated `$J=0$' to `the low-spin state' for
convenience. It does not means that the ground state is 
Eu${}^{3+} (J=0)$.  Instead, $E_{J=0}$ is defined by
the ground state energy per Eu atom $E_{\rm L}$ and 
$V(J=0) \equiv V_{\rm L}$.  Similarly, $E_{J_{\rm H}} = E_{\rm H}$
denotes the excitation energy per atom in the high-spin state 
and $V(J=J_{\rm H}) \equiv V_{\rm H}$.
%
%
\begin{figure}
\includegraphics[scale=1.0]{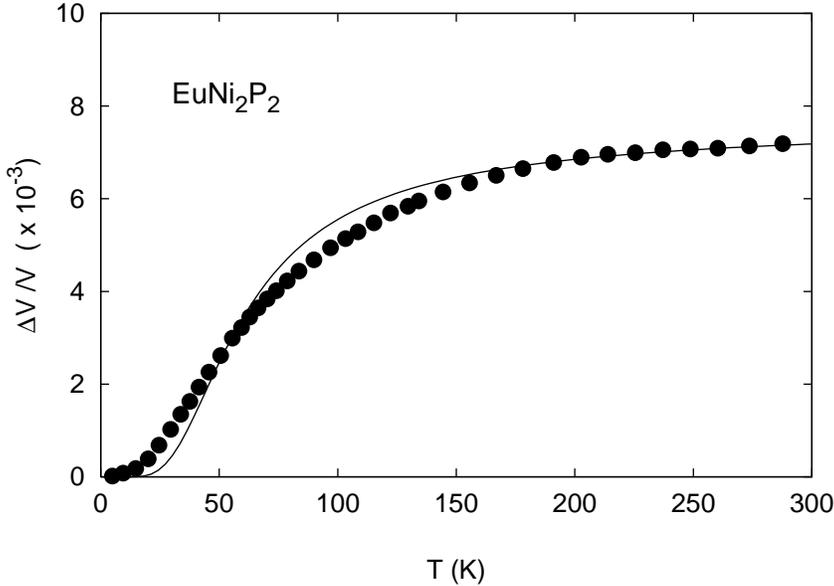}
\caption{
Temperature dependence of 4$f$ electron contribution to the thermal
 expansion ($\Delta V/V$). Closed circles: Experimental
 results~\cite{hiranaka13}, 
solid line: result of 2-state Weiss model for 
$\Delta E/k_{\rm B}=150$ K.
}
\label{figvol}
\end{figure}
%
%

After simple calculations of the
r.h.s. of Eq. (\ref{vol0}), we obtain the following expression for 
volume. 
\begin{eqnarray}
V = V_{\rm L} + v_{0}\frac{\displaystyle w \, e^{-\beta \Delta E}}
{\displaystyle 1 + w \, e^{-\beta \Delta E}} \, .
\label{vol1}
\end{eqnarray}
Here $v_{0}=V_{\rm H}-V_{\rm L}$, $w=2J_{\rm H}+1 \, (= 8)$, and
$\Delta E = E_{\rm H} - E_{\rm L}$ is the excitation energy from 
the low-spin state (L) to the high-spin state (H).
$T_{v} \equiv \Delta E/k_{\rm B}$ is interpreted as a valence
fluctuation temperature in the present system.  
The thermal expansion
$\Delta V/V_{\rm L} = (V-V_{L})/V_{L}$ is therefore given by
\begin{eqnarray}
\frac{\Delta V}{V_{\rm L}} = \frac{\Delta V(\infty)}{V_{\rm L}}
\frac{\displaystyle (1+w) \, e^{-\beta \Delta E}}
{\displaystyle 1 + w \, e^{-\beta \Delta E}} \, .
\label{vol2}
\end{eqnarray}
Here $\Delta V(\infty)/V_{\rm L}$ denotes the volume change in the
high-temperature limit. 

Figure 1  shows a numerical result of $\Delta V/V_{\rm L}$ compared
with the experimental data.  We adopted the experimental value
$\Delta V(\infty)/V_{\rm L} = 7.3 \times 10^{-3}$ at 300
K~\cite{hiranaka13} in the calculations.  
With use of a characteristic temperature 
$T_{v} =\Delta E/k_{\rm B}=150$ K,  
we find that the formula (\ref{vol2}) explains the overall feature of 
the experimental thermal expansion.  The deviation from the data at 
low temperatures is attributed to the fact that the model does not 
take into account the low-energy excitations associated with the 
heavy-fermion states.
%
%
\begin{figure}
\includegraphics[scale=1.0]{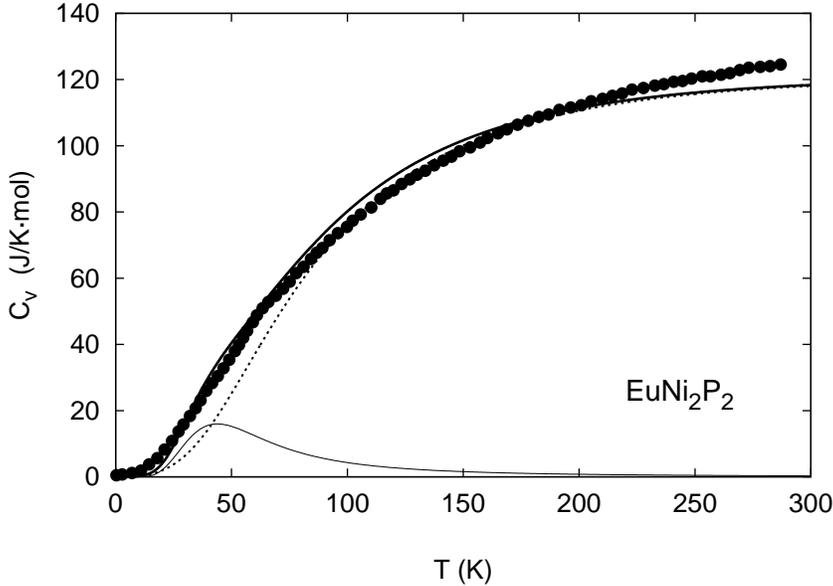}
\caption{
Specific heat in the present model (solid curve) and experimental data
 (closed circles)~\cite{hiranaka13}.
Thin solid curve shows the electronic contribution $C^{({\rm e})}_{v}$,
 which is calculated from Eq. (\ref{cev}) with use of the same parameter
 $\Delta E/k_{\rm B}=150$ K as in Fig. 1. Dashed curve is the lattice 
contribution calculated by the Debye model.  The Dulong-Petit parameter 
$A$ and the Debye temperature parameter $T_{\rm D}$ are chosen to be 
$A=15.9$ in unit of the gas constant and $T_{\rm D}=350$ K so that the 
experimental data around 50 K and 200 K are reproduced.    
}
\label{figcv}
\end{figure}
%
%

We obtain in the same way the average 4$f$ electron number as
\begin{eqnarray}
n_{f} = n_{f{\rm L}} + n_{0}\frac{\displaystyle w \, e^{-\beta \Delta E}}
{\displaystyle 1 + w \, e^{-\beta \Delta E}} \, .
\label{n4f}
\end{eqnarray}
Here $n_{0}=n_{f{\rm H}}-n_{f{\rm L}}$, $n_{f{\rm L}}$ ($n_{f{\rm H}}$)
being the electron number in the low-spin (high-spin) state.  
Note that $n_{f{\rm L}}$ is the average $f$ electron number per Eu 
atom in the heavy-fermion ground state ({\it i.e.}, 
$n_{f{\rm L}} \neq 6$).
Equation (\ref{n4f}) implies that there is a scaling relation 
between the volume
change $\Delta V/V_{\rm L}$ and that of 4$f$ electron 
number $\Delta n_{f} = n_{f}-n_{f{\rm L}}$:
\begin{eqnarray}
\frac{\Delta V}{V_{\rm L}} = \frac{v_{0}}{n_{0}V_{\rm L}} \, \Delta n_{f} .
\label{scalevn}
\end{eqnarray}
The relation is in agreement with the experimental 
fact~\cite{hiranaka13}.  According to
the M\"{o}ssbauer experiment, $n_{f}(T=0) \sim 6.5$ and 
$n_{f}(T=\infty) \sim 6.75$~\cite{nagara85}.  
Using Eq. (\ref{scalevn}) and these
values, we obtain $n_{f{\rm H}} \sim 6.8$, which is rather close to the
atomic value 7.0.

\subsection{Specific heat}

Similar scaling relation is found between the electronic contribution 
to the volume thermal expansion coefficient 
$\alpha^{({\rm e})}_{v}= V_{\rm L}^{-1} \partial V/\partial T$ and 
that to the heat capacity $C^{({\rm e})}_{v}$.  In fact, the
internal energy $E$ in the Weiss model is given by 
\begin{eqnarray}
E = E_{\rm L} + \Delta E \, \frac{\displaystyle w \, e^{-\beta \Delta E}}
{\displaystyle 1 + w \, e^{-\beta \Delta E}} \, .
\label{ener}
\end{eqnarray}
Thus we have a relation, 
\begin{eqnarray}
E - E_{\rm L} = \Delta E \frac{V_{\rm L}}{v_{0}} 
\frac{\Delta V}{V_{\rm L}} \, ,
\label{enern}
\end{eqnarray}
and find a scaling relation,
\begin{eqnarray}
\alpha^{({\rm e})}_{v} = 
\frac{1}{\Delta E} \frac{v_{0}}{V_{\rm L}} C^{({\rm e})}_{v} \, .
\label{alphav}
\end{eqnarray}
Here $C^{({\rm e})}_{v}$ is the electronic contribution in the present
model, being given by 
\begin{eqnarray}
C^{({\rm e})}_{v} =  \frac{\displaystyle w \, (\beta \Delta E)^{2} e^{-\beta \Delta E}}
{\displaystyle (1 + w \, e^{-\beta \Delta E})^{2}} \, .
\label{cev}
\end{eqnarray}

It should be noted however 
that the relation (\ref{alphav}) does not necessarily
mean that the Schottky-like anomaly is visible in the experimental data
of specific heat $C_{v}$, though the data of the volume expansion
coefficient show a clear anomaly around 40 K~\cite{hiranaka13}.
In order to see this point, we consider here a simple Debye model as a
lattice contribution $C^{({\rm l})}_{v}$ to the specific heat:
$C^{({\rm l})}_{v} = A D(T_{\rm D}/T)$. Here $A (\sim 15)$ is the
Dulong-Petit constant, $T_{\rm D}$ is the Debye temperature and $D(x)$
is the Debye function.  The total specific heat is given by 
$C_{v} = C^{({\rm e})}_{v} + C^{({\rm l})}_{v}$.
Figure 2 shows the calculated specific heat vs experimental data.
The present theory is consistent with the experimental data of the
specific heat, and we find that the anomalous contribution cannot be 
seen in the total $C_{v}$ because of a large lattice contribution of 
$C^{({\rm l})}_{v}$ and its steep slope in the temperature region 
between 40 K and 80 K~\cite{kake13}.

\subsection{Susceptibility}

The susceptibility in the 2-state Weiss model is obtained by adding
the Zeeman term to the Hamiltonian.  The magnetization 
$\langle M_{z} \rangle$ per Eu atom under the magnetic field $H$ is 
calculated from the following expression.
\begin{eqnarray}
\langle M_{z} \rangle = 
\frac{\displaystyle \sum_{JM} (M_{z})_{JMH} e^{-\beta E_{JMH}}}
{\displaystyle \sum_{JM} e^{-\beta E_{JMH}}}  .
\label{magz}
\end{eqnarray}
Here the `$J=0$' state is the heavy-fermion ground state under the
magnetic field $H$.  With use of the energy $E_{\rm L}$ and the
heavy-fermion ground-state susceptibility $\chi_{\rm L}$ which might 
be inversely proportional to the Kondo temperature, 
the energy under the magnetic field should be given by
\begin{eqnarray}
E_{JMH} = E_{\rm L} - \frac{1}{2} \chi_{\rm L} H^{2} .
\label{enerj0h}
\end{eqnarray}
The magnetization per atom in the low spin state is therefore given by 
$(M_{z})_{JMH} = - \partial E_{JMH}/ \partial H = \chi_{\rm L} H$.
On the other hand, the energy in the high spin state ($J=J_{\rm H}$) 
is given by 
\begin{eqnarray}
E_{JMH} = E_{J} - g_{J}MH - \frac{1}{2}\alpha_{JM} H^{2} .
\label{enerjmh}
\end{eqnarray}
Here $g_{J}$ is Land\'{e}'s $g$ factor, $\alpha_{JM}$ is a
phenomenological Van Vleck constant~\cite{takikawa10,nolting09}.
The magnetization is given by 
$(M_{z})_{JMH} = - \partial E_{JMH} / \partial H = 
g_{J}M + \alpha_{JM} H$.

From Eqs. (\ref{magz}), (\ref{enerj0h}), and (\ref{enerjmh}), 
we obtain the susceptibility.
\begin{eqnarray}
\chi = \chi_{\rm L} + \Big(\Delta \chi + \frac{p_{\rm H}^{2}}
{3k_{\rm B}T} \Big) \frac{\displaystyle w \, e^{-\beta \Delta E}}
{\displaystyle 1 + w \, e^{-\beta \Delta E}} \, .
\label{chi0}
\end{eqnarray}
Here $\Delta \chi = \chi_{\rm H} - \chi_{\rm L}$, and 
$\chi_{\rm H}$ is the susceptibility at $T=0$ in the high-spin 
state defined by $\sum_{M} \alpha_{J_{\rm H}M}/(2J_{\rm H}+1)$.  
$p_{\rm H}$ is the effective Bohr
magneton number in the high-spin state, {\it i.e.}, 
$p_{\rm H}^{2}=g_{J}^{2}J_{\rm H}(J_{\rm H}+1)$.
Taking the high-temperature limit, we obtain the effective 
Bohr magneton number $p_{\rm eff}$ as
\begin{eqnarray}
p_{\rm eff} = \sqrt{\frac{w}{1+w}} \, p_{\rm H} \, .
\label{peff}
\end{eqnarray}
Using the values $w=8$ and $p_{\rm H}=\sqrt{63}$, we obtain 
$p_{\rm eff} = 7.48$ $\mu_{\rm B}$, which is in good agreement with the
experimental value $7.4$ $\mu_{\rm B}$~\cite{hiranaka13}.  
%
%
\begin{figure}
\includegraphics[scale=1.0]{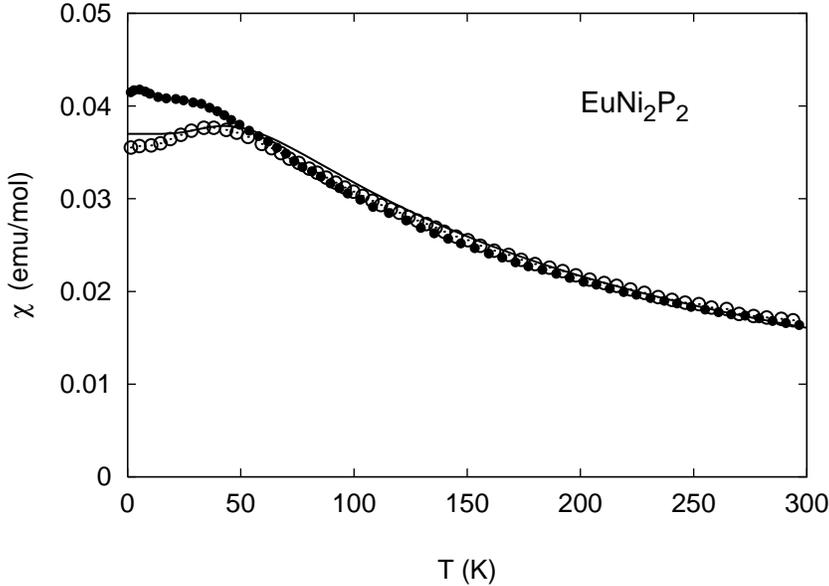}
\caption{
Temperature dependence of susceptibilities ($\chi$).
Open circles: experimental data~\cite{hiranaka13} for $H // [110]$,
closed circles: experimental data~\cite{hiranaka13} for $H // [001]$,
solid line: present result based on the 2-state Weiss model 
for $\chi_{\rm L} \approx 0.037$ emu/mol, $\Delta \chi =-0.044$ emu/mol,
 $\Theta=-120$ K, and $\Delta E/k_{\rm B}=150$ K 
(see Eq. (\ref{chimod})). 
}
\label{figchi}
\end{figure}
%
%

The susceptibility (\ref{chi0}) however does not explain an overall
temperature variation of the
experimental data.  We have to take into account the effect of 
polarization due to the RKKY-like magnetic interaction
via conduction electrons.  This produces an effective magnetic field
according to the mean-field picture and should produce the Weiss constant
$\Theta$ in the susceptibility when $J=J_{\rm H}$.  
We reach then the following susceptibility.
\begin{eqnarray}
\chi = \chi_{\rm L} + \Big(\Delta \chi + \frac{C_{\rm H}}
{T - \Theta} \Big) \frac{\displaystyle w \, e^{-\beta \Delta E}}
{\displaystyle 1 + w \, e^{-\beta \Delta E}} \, .
\label{chimod}
\end{eqnarray}
Here $C_{\rm H}=p_{\rm H}^{2}/3k_{\rm B}$ is the Curie constant 
in the high-spin state.  
It should be noted that the polarization effects
due to the magnetic field are negligible for the other
quantities because they are of order of $H^{2}$. 

Using the experimental values $\Theta=-120$ K and 
$\chi_{\rm L} \approx 0.037$ emu/mol~\cite{hiranaka13} and choosing
$\Delta \chi$ as 
$\Delta \chi =-0.044$ emu/mol, we can explain the global feature of
temperature dependence of EuNi${}_{2}$P${}_{2}$ as shown in Fig. 3. 
Note that we did not take into account the anisotropy due to 
tetragonal structure in the present analysis for simplicity.

\section{Summary}

We have shown in this paper that the phenomenological 2-state Weiss
model with an emphasis of local excitations explains an overall feature
of the volume expansion due to 4$f$ electrons as well as the scaling
relation between the 4$f$ electron contribution to the volume expansion 
and the average 4$f$ electron number.  
These results indicate that the strong temperature dependence of thermal
expansion found in this system is mainly determined by local excitations
associated with valence fluctuations.
We also predicted that there is another scaling
relation between the thermal expansion coefficient due to the 4$f$
electrons and the associated specific heat.  We have also shown that 
calculated effective Bohr
magneton number in the high-temperature limit is in good agreement with
the experimental value 7.4 $\mu_{\rm B}/$Eu.  For explanation of
temperature dependence of susceptibility, we need inter-site magnetic
interactions via conduction electrons.  
Modified susceptibility explains the temperature dependence of the 
experimental data, and indicates that there are three temperature
scales, the characteristic temperature 
$T_{v} = \Delta E/k_{\rm B} \ (\sim 150 {\rm K})$ 
for on-site
valence fluctuations, the Weiss constant $\Theta \ (\sim -120 {\rm K})$ 
associated with the
RKKY type interactions via conduction electrons, and the Kondo
temperature $T_{\rm K}$ 
for the formation of heavy-fermions.  
The results presented in this paper indicate that the heavy-fermion
ground state is collapsed by valence fluctuations which are 
characterized by
$T_{v}$. In this sense, the characteristic temperature $T_{v}$ plays 
an essential role in the physics of EuNi${}_{2}$P${}_{2}$. 
It has not yet been clarified theoretically whether or not the Kondo
temperature is well-defined in this system.

Although the present model is consistent with the heavy-fermion ground
state as well as the effective Bohr magneton number in the high
temperature limit which is directly related to the local excitations
of $4f{}^{7}$, it does not take into account the low energy excitations.
The temperature dependence associated with
the heavy-fermions is not described here. 
The deviation of $\Delta V/V$ from the
experimental data below 40 K in Fig. 1 should be attributed to this fact.
Too small a specific heat below 30 K found in Fig. 2 is due to the
neglect of the low energy excitations.  
Microscopic derivation of the phenomenological model and the inclusion
of the low-energy excitations are left for 
future investigations.

The present work is supported by a Grant-in-Aid for
Scientific Research (25400404).


%


\begin{thebibliography}{99}
\bibitem{kake13}
Needless to say, the $T$-linear specific heat due to heavy-fermions
	showing a weak temperature dependence is omitted in this
	analysis. But the conclusion derived here does not change even
	if we add such a term.
\end{thebibliography}
\end{document}